# A Variational Approach to the Structure and Thermodynamics of Linear Polyelectrolytes with Coulomb and Screened Coulomb Interactions

Bo Jönsson[1]

Physical Chemistry 2, Chemical Center,
Box 124, S-221 00 Lund, Sweden

Carsten Peterson[2] and Bo Söderberg[3]

Department of Theoretical Physics, University of Lund
Sölvegatan 14A, S-22362 Lund, Sweden

Abstract:

A variational approach, based on a discrete representation of the chain, is used to calculate free energy and conformational properties in polyelectrolytes. The true bond and Coulomb potentials are approximated by a trial isotropic harmonic energy containing monomer-monomer force constants as variational parameters. By a judicious choice of representation and the use of incremental matrix inversion, an efficient and fast-convergent iterative algorithm is constructed, that optimizes the free energy. The computational demand scales as $N^3$ rather than $N^4$ as expected in a more naive approach. The method has the additional advantage that in contrast to Monte Carlo calculations the entropy is easily computed. An analysis of the high and low temperature limits is given. Also, the variational formulation is shown to respect the appropriate virial identities. The accuracy of the approximations introduced are tested against Monte Carlo simulations for problem sizes ranging from $N = 20$ to 1024. Very good accuracy is obtained for chains with unscreened Coulomb interactions. The addition of salt is described through a screened Coulomb interaction, for which the accuracy in a certain parameter range turns out to be inferior to the unscreened case. The reason is that the isotropic Gaussian Ansatz becomes less efficient with shorter range interactions.

As a by-product a very efficient Monte Carlo algorithm was developed for comparisons, providing high statistics data for very large sizes – 2048 monomers. The Monte Carlo results are also used to examine scaling properties, based on low-$T$ approximations to end-end and monomer-monomer separations. It is argued that the former increases faster than linearly with the number of bonds.

[1] fk2boj@grosz.fkem2.lth.se
[2] carsten@thep.lu.se
[3] bs@thep.lu.se

# 1 Introduction

Polymers and polymer solutions play a profound role in our daily life, biologically and technologically. This is of course one reason for the intense theoretical studies of polymers, but they have also because of their very nature been a challenge to theoreticians. Much theoretical work has been done in order to obtain a general understanding of neutral polymers, either in melts or in solution [1, 2]. Polyelectrolytes, on the other hand, have been less thoroughly investigated, despite their importance in many technical applications – for example in glue production or in food industry, for promoting flocculation or in pulp drying [3, 4, 5]. The term polyelectrolyte is sometimes used as a collective name for any highly charged aggregate. However, here we will restrict the meaning to a flexible molecule with several or many charged or chargeable sites; poly-L-glutamic acid, polyamines and polysaccharides are some typical representatives.

The conformation of a flexible polyelectrolyte is a result of the competition between the covalent bonding forces, electrostatic interactions as well as more specific short ranged interactions. For example, poly-L-glutamic acid and several polysaccharides undergo helix to coil transition as a function of pH [6, 7]. This transition is obviously governed by electrostatic forces – similar structural transitions are seen for DNA [8]. Undoubtedly, both the polymer nature and the interaction between charged amino acids play an important role for the folding of a protein, as well as other solvent averaged forces. The present study should be seen as an attempt to approach the folding problem using what turns out to be a very powerful statistical mechanical variational technique. In the first step we will limit the study to linear polyelectrolytes in salt solution.

Variational methods are standard techniques in quantum mechanics, but less so in statistical mechanics, although variational principles were formulated many years ago [9]. One type of variational formulations starts off from an approximate free energy, which is optimized with respect to the particle density [10]. For polymers a more fundamental approach is possible, by introducing variational parameters directly into the appropriate Hamiltonian. In the past this route has been followed in a number of polymer studies [11, 12, 13] and it has recently been revitalized by several groups [14, 15]. To the best of our knowledge, all these calculations have been concerned with continous chains and only one or at most a few variational parameters have been optimized.

The present approach, of which some results were already published [16], is inspired by refs. [17, 18, 19]. It uses a discrete representation of the polymer, which not only allows us to investigate linear or cyclic polymers, but also polymers of arbitrary topology. Thus, e.g., a hyperbranched dendrite structure can easily be handled within this formalism. The number of variational parameters is proportional to $N^2$, where $N$ is the number of interacting units. In general, the variational approach is expected to be most accurate at high dimensions [17, 18]. Apparently, this has discouraged the community from pushing the approach for three-dimensional polymers into a numerical confrontation. Also, using the method in a naive way would give scaling behaviour like $N^4$, which would make the method less tractable for large sizes. We have also found empirically that such a naive implementation is plagued with bad convergence properties. In previous work [16] an algorithm was developed that lowers the computational costs to $N^3$ with controlled and nice convergence properties.

The high and low $T$ limits of the variational approach, which are accessible with analytical means, are derived in this paper. The low $T$ expansions for various quantities yield results that very well approximate what emerges from the corresponding expansions in the exact theory. The fact that the temperature of interest (room temperature) is fairly low on the temperature scale partly explains the success of initial numerical explorations of the variational approach [16] and motivates further studies.

Besides being more realistic, the discrete chain also offers the possibility of direct comparison with Monte Carlo (MC) simulations, thereby giving an important indication of the accuracy. The final output of the present variational calculation is of course the minimized free energy, but one also obtains a matrix with all possible monomer-monomer correlations within the chain. This matrix is the starting point for the calculation of end-end and monomer-monomer separations and different kinds of angular correlations.

MC simulations of flexible polyelectrolytes have only recently appeared in the literature and then limited to rather short chains [20, 21, 22, 23, 24, 25, 26, 27, 28], the longest chains being of the order of a few hundred monomers. So far simulations have dealt with both explicit representation of salt particles as well as implicit in the form of a screened Coulomb potential. The inclusion of short ranged interactions has also been studied as have the conformational changes associated with the titration of a flexible polyelectrolyte [27]. Most simulations have been carried out in the canonical ensemble, although the latter problem required the use of grand canonical MC simulations [26, 27]. The accuracy of the screened Coulomb potential has been investigated by several people and found to be an excellent approximation for not too high polyelectrolyte charge densities and in the absence of any multivalent ions [23, 29]. In order to obtain high statistics MC results a pivot algorithm and a recently developed hybrid scheme [30] were used. As a by-product these simulation results were also used to examine certain scaling properties, theoretically derived from an approximate low $T$ expression. In contrast to the case of rigid bonds we find that the end-end separations scale faster than linearly with N.

When confronting the variational approach with MC data the following results emerge in this work:

- The variational approach has the unique property that it directly yields the free energy $F$. This is in contrast to MC simulations, where F is only indirectly accessible through elaborate integrations. For N=20, where comparisons are inexpensive, the variational free energy nicely agrees with MC results.

- For unscreened Coulomb chains the success reported in ref. [16] survives to even larger systems (N=1024) for configurational properties like end-end correlations. Also for angular correlations and scaling behaviour the variational approach reproduces MC data very well.

- When including salt through Debye screened Coulomb potentials the performance of the method deteriorates somewhat on the quantitative level. Even if the qualitative configurational picture agrees with that from MC data, the actual numbers for e.g. end-end correlations could differ up to 50 %. We attribute this discrepancy partly to the inability of Gaussians to reproduce short range interactions.

One should also mention that other approximation schemes than the variational ones have been attempted in order to study polyelectrolyte conformations. In the mean field approximation it is only with the assumption of spherical symmetry that the mean field equations become tractable, but the results are not particularly encouraging [29]. In a cylindrical geometry, however, the mean field approximation behaves much more satisfactorily, but at the expense of large numerical efforts [31].

This paper is organized as follows: In sect. 2 the basic Coulomb models (screened and unscreened) are presented. The variational approach is presented in sect. 3. The high and low $T$ limits of the variational scheme are computed with analytical methods in sect. 4. A description of the MC method used in order to establish the quality of the variational method is found in Sect. 5. The results from confronting the variational approach with MC data are presented in sect. 6. Finally in sect. 7 a brief summary and outlook is given. Most of the detailed derivations are found in appendices – generics about the variational method (A), variational energies for a polyelectrolyte (B), virial identities (C), high and low $T$ expansions (D) and zero temperature scaling properties (E). The disposition of the material into bulk text and appendices is such that the approach and results can be fully understood without reading the appendices.

## 2 The Model

### 2.1 The Unscreened Coulomb Chain

In this model we consider a polyelectrolyte at infinite dilution and without any added salt. The polyelectrolyte counterions are thus neglected and the only electrostatic interactions are between the charged monomers. More explicitly, the polymer chain consists of $N$ point charges connected by harmonic oscillator ("Gaussian") bonds. The potential energy for a chain then takes the form

$$\tilde{E} = \tilde{E}_G + \tilde{E}_C = \frac{k}{2} \sum_{i=1}^{N-1} |\tilde{\mathbf{x}}_{i,i+1}|^2 + \frac{q^2}{4\pi\epsilon_r\epsilon_0} \sum_{i<j} \frac{1}{|\tilde{\mathbf{x}}_{ij}|} \quad (1)$$

Here, $\tilde{\mathbf{x}}_i$ the position of the $i$th charge, and

$$\tilde{\mathbf{x}}_{ij} = \tilde{\mathbf{x}}_i - \tilde{\mathbf{x}}_j \quad (2)$$

while $q$ is the monomer charge and $\epsilon_r\epsilon_0$ is the dielectric permittivity of the medium. We use the tilde notation $\tilde{E}$, $\tilde{\mathbf{x}}_i$, etc. for physical quantities in conventional units, and reserve $E$, $\mathbf{x}_i$, etc. for dimensionless ones, which will be used in the theoretical formalism below.

The force constant can be reexpressed in terms of the $N = 2$ equilibrium distance $r_0$, given by

$$k = \frac{q^2}{4\pi\epsilon_r\epsilon_0} \frac{1}{r_0^3} \quad (3)$$

In terms of the dimensionless coordinates $\mathbf{x}_i$, defined by

$$\tilde{\mathbf{x}}_i = r_0 \mathbf{x}_i \quad (4)$$

the energy takes the form

$$\tilde{E} = kr_o^2 \left( \frac{1}{2} \sum_i |\mathbf{x}_{i,i+1}|^2 + \sum_{i<j} \frac{1}{|\mathbf{x}_{ij}|} \right) \quad (5)$$

We will consider the system at a finite temperature $\tilde{T}$, which can be similarly rescaled,

$$T = \frac{k_B \tilde{T}}{kr_0^2} \quad (6)$$

with $k_B$ the Boltzmann constant. One then obtains for the Boltzmann exponent the simple expression

$$\frac{\tilde{E}}{k_B \tilde{T}} = \frac{E}{T} = \frac{1}{T} \left( \frac{1}{2} \sum_i |\mathbf{x}_{i,i+1}|^2 + \sum_{i<j} \frac{1}{|\mathbf{x}_{ij}|} \right) \tag{7}$$

In other words, a polyelectrolyte at infinite dilution represents a two-parameter model where $T$ and the number of monomers $N$ are the only two non-trivial parameters of the system.

Unless otherwise stated the following parameter values will be used throughout the paper; $\tilde{T}$=298K, $\epsilon_r$=78.3 and $r_0$=6Å. Obviously, $E$ depends only on the relative positions; the global center-of-mass position variable will have to be excluded from integrations over the coordinate space.

The Gaussian and Coulomb energies are subject to a *virial identity* (see Appendix C), which in dimensionless units reads

$$2\langle E_G \rangle - \langle E_C \rangle = 3(N-1)T \tag{8}$$

Eq. (8) is a useful relation for checking the correctness and convergence behaviour in MC simulations and we find that it is in general obeyed to 0.3% or better.

## 2.2 The Screened Coulomb Chain

To treat a single polyelectrolyte in a solution at finite salt concentration becomes very costly in a MC simulation, since for reasonable salt concentrations the number of salt ions easily becomes prohibitively large, much larger than the number of polyelectrolyte monomers. The usual way to avoid this problem is to preaverage the degrees of freedom of the simple salt ions for some fixed configuration of the polyelectrolyte [32], thus defining salt averaged effective potentials – this is the basis of the electrostatic contribution in the classical DLVO potential [33]. In this way we may derive a screened Coulomb potential from a linearisation of the Poisson-Boltzmann equation for the salt. Eq. (1) is then replaced by

$$\tilde{E} = \tilde{E}_G + \tilde{E}_C = \frac{k}{2} \sum_i |\tilde{\mathbf{x}}_{i,i+1}|^2 + \frac{q^2}{4\pi\epsilon_r\epsilon_0} \sum_{i<j} \frac{e^{-\tilde{\kappa}|\tilde{\mathbf{x}}_{ij}|}}{|\tilde{\mathbf{x}}_{ij}|} \tag{9}$$

where $\tilde{\kappa}$ is the Debye screening length for a 1:1 salt defined as

$$\tilde{\kappa} = q \sqrt{\frac{2 N_A c_s}{\epsilon_r \epsilon_0 k_B T}} \tag{10}$$

In eq. (10) $c_s$ is the salt concentration in molars (M) and $N_A$ is the Avogadro's number. The Boltzmann factor will for the screened Coulomb potential contain the inverse dimensionless Debye screening length, $\kappa = r_0 \tilde{\kappa}$ as an additional parameter, and with the parameter values given above we have for $c_s = 0.01$ M, 0.1 M and 1.0 M the $\kappa$-values 0.1992, 0.6300 and 1.992 respectively. Then eq. (7) is modified into

$$\frac{E}{T} = \frac{1}{T} \left( \frac{1}{2} \sum_i |\mathbf{x}_{i,i+1}|^2 + \sum_{i<j} \frac{e^{-\kappa |\mathbf{x}_{ij}|}}{|\mathbf{x}_{ij}|} \right) \tag{11}$$

The virial identity (see Appendix C) now takes the form

$$2\langle E_G\rangle - \langle E_C\rangle - \kappa \langle \sum_{i<j} e^{-\kappa|\mathbf{x}_{ij}|}\rangle = 3(N-1)T \qquad (12)$$

## 2.3 Relative coordinates

In the remainder of this paper, relative coordinates will be used; instead of the absolute monomer positions $\mathbf{x}_i$, the *bond vectors* $\mathbf{r}_i$,

$$\mathbf{r}_i \equiv \mathbf{x}_{i+1} - \mathbf{x}_i, \quad i = 1,\ldots, N-1 \qquad (13)$$

will be used as the fundamental variables. In this way complications due to the translational zero-mode are avoided; in addition the convergence of the algorithm is considerably speeded up, especially at high temperatures.

The energy of the screened Coulomb chain will then take the following form:

$$E(\mathbf{r}) = E_G + E_C = \frac{1}{2}\sum_{i=1}^{N-1}\mathbf{r}_i^2 + \sum_\sigma \frac{e^{-\kappa r_\sigma}}{r_\sigma} \qquad (14)$$

where $\sigma$ runs over contiguous non-nil sub-chains, with

$$\mathbf{r}_\sigma \equiv \sum_{i\in\sigma}\mathbf{r}_i \qquad (15)$$

corresponding to the distance vector between the endpoints of the subchain. The unscreened chain results for $\kappa = 0$.

# 3 The Variational Approach

## 3.1 The Gaussian Ansatz

In refs. [18, 19, 16] the variational method of refs. [9, 17] (see Appendix A for a generic description) was revisited in the context of discrete chains of polyelectrolytes. The approach is based on an effective energy Ansatz $E_V$, given by

$$E_V/T = \frac{1}{2}\sum_{ij} G^{-1}_{ij}(\mathbf{r}_i - \mathbf{a}_i)\cdot(\mathbf{r}_j - \mathbf{a}_j) \qquad (16)$$

where $\mathbf{a}_i$ defines an average bond vector, around which Gaussian fluctuations are given by the symmetric positive-definite correlation matrix $G_{ij}$, the matrix inverse of which appears in the energy.

Using this effective energy, the exact free energy $F = -T\log Z$ of the polymer is approximated from above [9] by the variational one

$$\hat{F} = F_V + \langle E - E_V\rangle_V \geq F \qquad (17)$$

where $F_V = -T \log Z_V$, and $\langle \rangle_V$ refers to averages with respect to the trial Boltzmann distribution $\exp(-E_V/T)$.

The parameters $G_{ij}$ and $\mathbf{a}_i$ are to be determined such that the variational free energy $\hat{F}$ is minimized; we note that the number of variational parameters increases with $N$ like $N^2$. The resulting effective Boltzmann distribution is then used to approximate expectation values $\langle f \rangle$ by effective ones $\langle f \rangle_V$. Thus, we have e.g.

$$\langle \mathbf{r}_i \rangle_V = \mathbf{a}_i \qquad (18)$$
$$\langle \mathbf{r}_i \cdot \mathbf{r}_j \rangle_V = \mathbf{a}_i \cdot \mathbf{a}_j + 3 G_{ij} \qquad (19)$$

For the polyelectrolyte systems treated in this study we will at high temperatures find a unique variational solution, characterized by $\mathbf{a}_i = 0$; this defines a *purely fluctuating* solution. At low temperatures we find in addition a *rigid* solution with aligned $\mathbf{a}_i \neq 0$. The latter is due to spontaneous symmetry-breaking; it ceases to exist at high temperatures, but will at low enough temperatures have the lower free energy. As discussed below, the rigid solution can typically be disregarded at normal temperatures.

For potentials more singular than $1/r^2$, $\langle E \rangle_V$ will be divergent, and the approach breaks down. However, such potentials are not physical and we do not consider this limitation of the approach a serious one.

A non-trivial result of the scaling properties of the effective energy is that the virial identity, eqs. (8,12), will be respected by the above variational approach (see Appendix C).

## 3.2 Using Local Fluctuation Amplitudes

The minimization of $\hat{F}$ with respect to $G_{ij}$ and $\mathbf{a}_i$ gives rise to a set of matrix equations to be solved iteratively. These are considerably simplified, and the symmetry and positivity constraints on $G_{ij}$ are automatic, if $G_{ij}$ is expressed as the product of a matrix and its transpose:

$$G_{ij} = \sum_{\mu=1}^{N-1} z_{i\mu} z_{j\mu} = \mathbf{z}_i \cdot \mathbf{z}_j \qquad (20)$$

The interpretation of the local parameter $\mathbf{z}_i$ is simple – it is a fluctuation amplitude for the $i$th bond vector $\mathbf{r}_i$. We can write

$$\mathbf{r}_i = \mathbf{a}_i + \sum_\mu z_{i\mu} \mathbf{J}_\mu \qquad (21)$$

where each component of $\mathbf{J}_\mu \in \mathcal{R}^3$ is an independent Gaussian noise variable of unit variance.

Similarly, we have for a subchain

$$\mathbf{r}_\sigma = \sum_{i \in \sigma} \mathbf{r}_i \equiv \mathbf{a}_\sigma + \sum_\mu z_{\sigma\mu} \mathbf{J}_\mu, \qquad (22)$$

where $\mathbf{a}_\sigma = \sum_{i \in \sigma} \mathbf{a}_i$ and $\mathbf{z}_\sigma = \sum_{i \in \sigma} \mathbf{z}_i$. Thus, the noise amplitudes are additive.

The matrix inverse of $G$ can similarly be decomposed:

$$G^{-1}_{ij} = \mathbf{w}_i \cdot \mathbf{w}_j \qquad (23)$$

where $w_{i\mu}$ is the (transposed) matrix inverse of $z_{i\mu}$:

$$\mathbf{z}_i \cdot \mathbf{w}_j = \delta_{ij} \qquad (24)$$

Note that $\mathbf{z}_i$, $\mathbf{w}_i$ and $\mathbf{z}_\sigma$ are vectors, not in $R^3$, but in $R^{N-1}$.

The equations for a local extremum of $\hat{F}(\mathbf{a}, \mathbf{z})$ are obtained by differentiation with respect to $\mathbf{z}_i$ and $\mathbf{a}_i$,

$$\frac{\partial \hat{F}}{\partial \mathbf{z}_i} = 0 \ , \ \frac{\partial \hat{F}}{\partial \mathbf{a}_i} = 0 \qquad (25)$$

## 3.3 The Unscreened Coulomb Chain

In terms of $\mathbf{a}_i$ and $\mathbf{z}_i$, the variational free energy for the pure Coulomb chain becomes, ignoring trivial additive constants (see Appendix B):

$$\hat{F} = -3T \log \det z + \frac{1}{2} \sum_i (3\mathbf{z}_i^2 + \mathbf{a}_i^2) + \sum_\sigma \frac{1}{a_\sigma} \operatorname{erf}\left(\frac{a_\sigma}{\sqrt{2}\, z_\sigma}\right) \qquad (26)$$

The equations for a minimum will be

$$\frac{\partial \hat{F}}{\partial \mathbf{z}_i} = -3T\mathbf{w}_i + 3\mathbf{z}_i - \sqrt{\frac{2}{\pi}} \sum_{\sigma \ni i} \frac{\mathbf{z}_\sigma}{z_\sigma^3} \exp\left(-\frac{a_\sigma^2}{2 z_\sigma^2}\right) = 0 \qquad (27)$$

$$\frac{\partial \hat{F}}{\partial \mathbf{a}_i} = \mathbf{a}_i - \sum_{\sigma \ni i} \frac{\mathbf{a}_\sigma}{a_\sigma^3}\left[\sqrt{\frac{2}{\pi}}\frac{a_\sigma}{z_\sigma} - \operatorname{erf}\left(\frac{a_\sigma}{\sqrt{2}\, z_\sigma}\right)\right] = 0 \qquad (28)$$

where the reciprocal vector $\mathbf{w}_i$ is defined by eq. (24).

These equations allow a purely fluctuating solution with $\mathbf{a}_i = 0$. Setting $\mathbf{a}_i = 0$ the variational free energy simplifies to

$$\hat{F} = -3T \log \det z + \frac{3}{2} \sum_i \mathbf{z}_i^2 + \sqrt{\frac{2}{\pi}} \sum_\sigma \frac{1}{z_\sigma} \qquad (29)$$

which looks very much like the energy of an $(N-1)$-dimensional Coulomb chain with bonds $\mathbf{z}_i$, but with an extra *entropy* term (the first) preventing alignment of the ground state. The $\mathbf{z}$ derivatives become

$$\frac{\partial \hat{F}}{\partial \mathbf{z}_i} = -3T\mathbf{w}_i + 3\mathbf{z}_i - \sqrt{\frac{2}{\pi}} \sum_{\sigma \ni i} \frac{\mathbf{z}_\sigma}{z_\sigma^3} = 0 \qquad (30)$$

## 3.4 The Screened Coulomb Chain

In the case of Debye screening the expression for $\hat{F}$ is modified to

$$\hat{F} = -3T \log \det z + \frac{1}{2} \sum_i (3\mathbf{z}_i^2 + \mathbf{a}_i^2) \tag{31}$$

$$+ \sum_\sigma \frac{1}{2a_\sigma} \exp\left(-\frac{a_\sigma^2}{2z_\sigma^2}\right) \left\{ \Psi\left(\kappa z_\sigma - \frac{a_\sigma}{z_\sigma}\right) - \Psi\left(\kappa z_\sigma + \frac{a_\sigma}{z_\sigma}\right) \right\}$$

where

$$\Psi(x) \equiv \exp(x^2/2) \, \mathrm{erfc}(x/\sqrt{2}) \tag{32}$$

The corresponding derivatives (cf. eq. (27)) take the form

$$\frac{\partial \hat{F}}{\partial \mathbf{z}_i} = -3T\mathbf{w}_i + 3\mathbf{z}_i \tag{33}$$

$$- \sum_{\sigma \ni i} \frac{\kappa^2 \mathbf{z}_\sigma}{2a_\sigma} \exp\left(-\frac{a_\sigma^2}{2z_\sigma^2}\right) \left\{ \Psi\left(\kappa z_\sigma - \frac{a_\sigma}{z_\sigma}\right) - \Psi\left(\kappa z_\sigma + \frac{a_\sigma}{z_\sigma}\right) - \sqrt{\frac{2}{\pi}} \frac{a_\sigma}{\kappa^2 z_\sigma^3} \right\} = 0$$

$$\frac{\partial \hat{F}}{\partial \mathbf{a}_i} = \mathbf{a}_i \tag{34}$$

$$- \sum_{\sigma \ni i} \frac{\kappa \mathbf{a}_\sigma}{2a_\sigma^2} \exp\left(-\frac{a_\sigma^2}{2z_\sigma^2}\right) \left\{ \Psi\left(\kappa z_\sigma - \frac{a_\sigma}{z_\sigma}\right) + \Psi\left(\kappa z_\sigma + \frac{a_\sigma}{z_\sigma}\right) - \sqrt{\frac{2}{\pi}} \frac{2}{\kappa z_\sigma} \right\} = 0$$

Also here, a purely fluctuating solution is allowed. Setting $\mathbf{a}_i = 0$ the variational free energy reduces to

$$\hat{F} = -3T \log \det z + \frac{3}{2} \sum_i \mathbf{z}_i^2 + \sum_\sigma \left\{ \sqrt{\frac{2}{\pi}} \frac{1}{z_\sigma} - \kappa \Psi(\kappa z_\sigma) \right\} \tag{35}$$

The $\mathbf{z}$ derivatives will be

$$\frac{\partial \hat{F}}{\partial \mathbf{z}_i} = -3T\mathbf{w}_i + 3\mathbf{z}_i - \sum_{\sigma \ni i} \frac{\mathbf{z}_\sigma}{z_\sigma^3} \left\{ \sqrt{\frac{2}{\pi}} (1 - \kappa^2 z_\sigma^2) + \kappa^3 z_\sigma^3 \Psi(\kappa z_\sigma) \right\} = 0 \tag{36}$$

## 3.5 Implementation

Due to the use of relative coordinates and of local noise amplitudes, a simple gradient descent method with a large step-size $\epsilon$ can be used, that gives fast convergence to a solution of eqs. (25)

$$\Delta \mathbf{z}_i = -\epsilon_z \frac{\partial \hat{F}}{\partial \mathbf{z}_i}, \quad \Delta \mathbf{a}_i = -\epsilon_a \frac{\partial \hat{F}}{\partial \mathbf{a}_i}, \tag{37}$$

Further speed is gained by updating the reciprocal variables $\mathbf{w}_i$ using incremental matrix inversion [16] – the increment in $\mathbf{w}_j$ due to $\Delta \mathbf{z}_i$ is given (exactly) by

$$\Delta \mathbf{w}_j = -\frac{\mathbf{w}_i(\mathbf{w}_j \cdot \Delta \mathbf{z}_i)}{1 + \mathbf{w}_i \cdot \Delta \mathbf{z}_i} \tag{38}$$

to be applied in parallel for $j$ for fixed $i$. As a by-product the denominator $(1 + \mathbf{w}_i \cdot \Delta \mathbf{z}_i)$ gives the multiplicative change in the determinant det $z$, needed to keep track of $\hat{F}$.

In order to maintain a reasonable numerical precision, the erf-related functions needed in the process are evaluated using carefully defined Taylor expansions or asymptotic expansions, depending on the size of the argument.

As discussed above, the equations for a minimum are consistent with a purely fluctuating solution $\mathbf{a}_i = 0$. Such a solution does indeed exist at all $T$; furthermore it is the only solution for high enough $T$. It turns out that for realistic choices of $T$ one is in the region where this solution gives rise to good results (see figs. 1, 2 below). The additional solution with $\mathbf{a}_i \neq 0$ appearing at low $T$ is a symmetry-broken solution; to be realistic, such a solution should show an anisotropy also in the fluctuations, i.e. different amplitudes $\mathbf{z}_i$ for the fluctuations parallel and transverse to the direction defined by (the aligned) $\mathbf{a}_i$. This requires a more general Ansatz than eq. (16); theoretically, this will produce better low-$T$ solutions, but for the Coulomb chain it leads to equations containing functions that are difficult to evaluate numerically, so we will not use this possibility in this paper. The incomplete symmetry-breaking partly explains the tendency for the $\mathbf{a} \neq 0$ solutions to produce inferior solutions.

For the above reasons, and for reasons of continuity, we will in the numerical explorations use the $\mathbf{a} = 0$ solutions, where not otherwise stated. This also implies faster performance since only the $\mathbf{z}_i$-variables, with simplified updating equations, are needed.

The complete algorithm will look as follows:

1. Initialize $\mathbf{z}_i$ (and $\mathbf{a}_i$ if present) randomly, suitably in the neighborhood of a truncated high- or low-$T$ series solution.

2. For each $i$:

   - Update $\mathbf{z}_i$ (and $\mathbf{a}_i$) according to eqs. (37) with suitable step-sizes.
   - Correct all $\mathbf{w}_j$ according to eq. (38).

3. Check if converged; if not, go to 2.

4. Extract $\mathbf{a}_i$ and $G_{ij} = \mathbf{z}_i \cdot \mathbf{z}_j$, and compute variational averages of interest.

Typical step-sizes are $\epsilon_z \approx 1/6$ and $\epsilon_a \approx 1/2$. The convergence check is done based on the rate of change and on the virial identity. The number of computations in each iteration step for this procedure is proportional to $N^3$. The number of iterations required for convergence is a slowly growing function $g(N)$. In total, thus, the execution time of the algorithm grows with $N$ as $N^3 g(N)$. In terms of CPU requirement convergence of a $N = 40$ chain ($\mathbf{a}_i = 0$) requires 3 seconds on a DEC Alpha workstation.

# 4 High and Low T Results – Analytical Considerations

Before embarking on a numerical evaluation of the variational approach with comparisons to MC results, it is interesting to see what can be gained from studying the high and low $T$ limits, where analytical methods can be used.

At the energy minimum, prevailing at $T = 0$, the polyelectrolyte will form a straight line. When the temperature is increased there will be a competition between the entropy and the repulsive Coulomb forces, and as $T \to \infty$, the chain becomes Brownian, and the elongated or ordered structure is gone altogether. The temperature range of the transition from ordered to disordered structure is $N$-dependent. As $N$ increases at fixed $T$, the Coulomb force becomes relatively more important and the system effectively behaves as if $T$ decreased: the polymer configuration becomes increasingly aligned.

In the variational approach, as discussed above, the high temperature regime is characterized by a purely fluctuating solution, reflecting the Brownian nature of the chain at $T \to \infty$. Such a solution survives as a local minimum also at lower $T$, where however also a rigid solution exists. Below a certain critical temperature $T_c$, the latter gives the global minimum, indicating a first order phase transition. This is probably an artefact of the variational approach - in the MC simulations the system shows no evidence of possessing a phase transition (see section 5). The rigid solution mirrors the ordered elongated structure of the polymer at low $T$.

With these qualitative arguments in mind, we turn to a more detailed investigation of the behaviour of the polyelectrolyte in the high and low $T$ limits, together with an evaluation of the corresponding variational results. The unscreened and screened cases will be treated separately.

## 4.1 The Unscreened Coulomb case

### 4.1.1 High Temperature

In the high $T$ limit, the variational results can be expanded in $1/T$ (see Appendix D). Thus, for the expectation value of the Gaussian energy $E_G$, the first two terms of the expansion yield,

$$\langle E_G \rangle = 3(N - 1)T/2 + 1/\sqrt{2\pi T} \sum_{k=1}^{N-1} \frac{N - k}{\sqrt{k}} \tag{39}$$

which agrees with the exact result to the order shown; the first discrepancy occurs in the $O(T^{-2})$ term, as is in fact true for any quadratic expectation value $\langle \mathbf{r}_i \mathbf{r}_j \rangle$. By the virial identity, this also holds for the Coulomb energy $\langle E_C \rangle$ and for the total energy as well.

For the individual bond lengths, the high $T$ result can be written

$$\langle \mathbf{r}_i^2 \rangle \approx 3T + 4\sqrt{2/\pi T} \left( \sqrt{i} + \sqrt{N - i} - \sqrt{N} \right) \tag{40}$$

where the last term is obtained from a continuum approximation, valid for large $N$.

### 4.1.2 Low Temperature

In the low $T$ limit (see Appendix D), the exact result for the total internal energy, expanded in powers of $T$, is,

$$\langle E \rangle = E_0 + (3N - 5)T/2 + O(T^2) \tag{41}$$

where $E_0$ is the minimum energy at $T = 0$ and $3N - 5$ is the number of degrees of freedom, modified for the spherical symmetry ($\rightarrow -2$). It turns out that with the above first order low $T$ correction, the energy, and thus also $E_G$, $E_C$ and $r_{mm}$ (see below), are quite well approximated for the sizes and temperatures considered in this paper.

At low $T$, the variational free energy is minimized by the rigid solution, for which the corresponding expansion is,

$$\langle E \rangle_V = E_0 + 3(N - 1)T/2 + O(T^2) \tag{42}$$

where the first term is the same as in eq. (41), while the second term is qualitatively correct for large $N$.

Yet another low temperature expansion results from the purely fluctuating variational solution and it gives,

$$\langle E \rangle_V = (6/\pi)^{1/3} E_0 + 3(N - 2)T/2 + O(T^2) \tag{43}$$

which shows a 24% discrepancy in the first term. The same factor, $(6/\pi)^{1/3}$, results for any quadratic expectation value and for any single term in $\langle E_C \rangle$ in this limit. For r.m.s. distances, like the monomer-monomer distance $r_{mm}$,

$$r_{mm} = \left\langle \frac{1}{N - 1} \sum_i \mathbf{r}_i^2 \right\rangle^{1/2} \tag{44}$$

or the end-end distance $r_{ee}$,

$$r_{ee} = \left\langle \left( \sum_i \mathbf{r}_i \right)^2 \right\rangle^{1/2} \tag{45}$$

this corresponds to an error of 11%.

With the variational approach thus satisfactory in both temperature limits one can hope to find it a reasonable approximation also at finite temperatures, as will indeed be borne out in section 5.

### 4.1.3 Zero Temperature

Having low $T$ expansions under control in terms of the $T = 0$ configurational properties, the latter remain to be calculated. They are given by the minimum energy configuration, which is aligned,

$$\mathbf{r}_i = b_i \hat{\mathbf{n}} \tag{46}$$

and unique (up to global translations and rotations).

The bond lengths $b_i > 0$ satisfy the equation

$$b_i = \sum_{\sigma \ni i} \frac{1}{b_\sigma^2} \tag{47}$$

where $b_\sigma = \sum_{j \in \sigma} b_j$ is the length of the subchain $\sigma$. This equation cannot be solved analytically (except for very small $N$), but a fair large-$N$ approximation can be obtained (see Appendix E for details). This gives for a distinct monomer-monomer bond the result

$$\langle \mathbf{r}_i^2 \rangle_{T=0}^{1/2} \equiv b_i \approx \left[ \log \left( \text{const} \, \frac{i(N-i)}{N} \right) \right]^{1/3} \tag{48}$$

As a consequence, we find that at zero temperature the average bond-length should scale logarithmically with $N$,

$$r_{mm} \propto (log N)^{1/3} \tag{49}$$

For the end-end separation this implies

$$r_{ee} \propto N (\log N)^{1/3} \tag{50}$$

This result is interesting, since it predicts a scaling faster than $N$; the extra logarithmic factor comes from the stretching of the harmonic bonds.

The results above might seem as rather academic and of little practical impact. However, the low temperature expressions turn out to be quite accurate when compared with MC results. In other words ordinary room temperature and aqueous solution corresponds to a surprisingly "low" temperature. This indicates that the low temperature expansion might be a good starting point for further work in polyelectrolyte solutions.

## 4.2 The Screened Coulomb case

At high $T$, an analysis similar to the one carried out for the pure Coulomb chain, can be done for the screened chain (see Appendix D). Also there, the variational approximations to quadratic expectation values turn out correct to next-to-leading order; the same holds for $E$ and $E_G$, while $E_C$ (which is one order down) is correct to leading order.

Also at low $T$, the results remain essentially the same for the screened chain. Thus, the rigid solution gives correct energies to lowest order in $T$, while the purely fluctuating solution does not.

# 5 Monte Carlo Simulation Techniques

For the numerical evaluation of the variational approach, the results were compared to those from MC simulations, which were performed in the canonical ensemble with the traditional Metropolis algorithm [34]. For short chains this is a straightforward procedure, but for linear chains consisting of more than about 100 monomers convergence problems appear. Typically for a chain of 40 monomers four million moves/monomer were required in order for the statistical fluctuations in the end-end separation to be less than one per cent. The energy terms and local conformational properties like monomer-monomer separations converged much faster. The addition of salt, i.e. the use of a screened potential, improves the convergence characteristics, but on the other hand its evaluation is more time-consuming than the pure inverse square root. Careful coding, with table look-ups for the inverse square root routine and, in particular, for the screened Coulomb potential, turned out to be more rewarding, reducing the computation time by almost a factor of three [35].

In order to treat longer chains with reasonable statistics we are forced to use more efficient algorithms like the pivot algorithm, first described in refs. [36, 37], with a high efficiency for linear chains on a lattice with short range interactions. Recently it has also been used successfully [38] for off-lattice simulations of a single polymer chain. It could be argued that the pivot algorithm should be even more efficient for chains with long range repulsive interactions like a charged polymer. The form of the pivot procedure used in this work can be described as a two step process, consisting of a random translation followed by a random rotation or vice versa. For a polymer chain with fixed bond lengths only random rotations will be used. The procedure is as follows: choose a monomer $i$ and apply the same random translation to monomers $i+1$ to $N$. Then choose an axis at random and perform a random rotation of monomers $i+1$ to $N$ around this axis. Evaluate the interaction energy between monomers 1 to $i$, and $i+1$ to $N$. This is a quadratic process in contrast to the single move algorithm, which only requires the evaluation of $N$ pair interactions/move. Finally a Metropolis energy criterion is used to test for rejection or acceptance of the new configuration. We find that with a maximal random displacement of the order of 5-10 Å and a maximal random rotation of $\pi$, we reject approximately 50 % of the attempted moves. Typically, we generate $10^3$-$10^4$ passes (one pass = one attempted move/monomer) resulting in a statistical uncertainty in the end-end separation of approximately one per cent. The uncertainty in the average monomer-monomer separation and in the Coulomb and Gaussian energies is much less. Local averages, however, like the $i$th bond length, may have larger uncertainties, something that is also discussed by Madras and Sokal [37]. The pivot algorithm seems to be superior to the traditional single monomer procedure described initially already for chains with $N > 20$. We also have found that restricting the procedure to translational moves still makes it superior to the traditional algorithm. The pivot algorithm makes it feasible to simulate chains with more than one thousand interacting monomers. The major drawback with the pivot algorithm seems to be its limitations to linear chains or at least chains with simple topologies.

Without excessive fine tuning of the translational and rotational displacement parameters, we find that the computational cost grows as $N^3$. This power results from $N^2$ for each sweep of monomer moves, and an additional factor $N$ from autocorrelations in quantities like end-end separations.

Another efficient algorithm has recently been developed in ref. [30]. By identifying the slow modes in a Fourier analysis, one is able to use different random step lengths for different modes. This technique seems to be as efficient as the pivot algorithm and we have used it to check the accuracy of our simulations. For all cases investigated we obtain, within the statistical uncertainties, identical averages. The same is true for shorter chains where we also can use the original single monomer algorithm as a further test.

# 6 Numerical Results

The superiority of the purely fluctuating variational solution over the rigid one, as discussed in section 5, is illustrated in figs. 1 and 2, where the variational results for $r_{ee}$ and $r_{mm}$ are compared to MC data for $N = 20$ and 80.

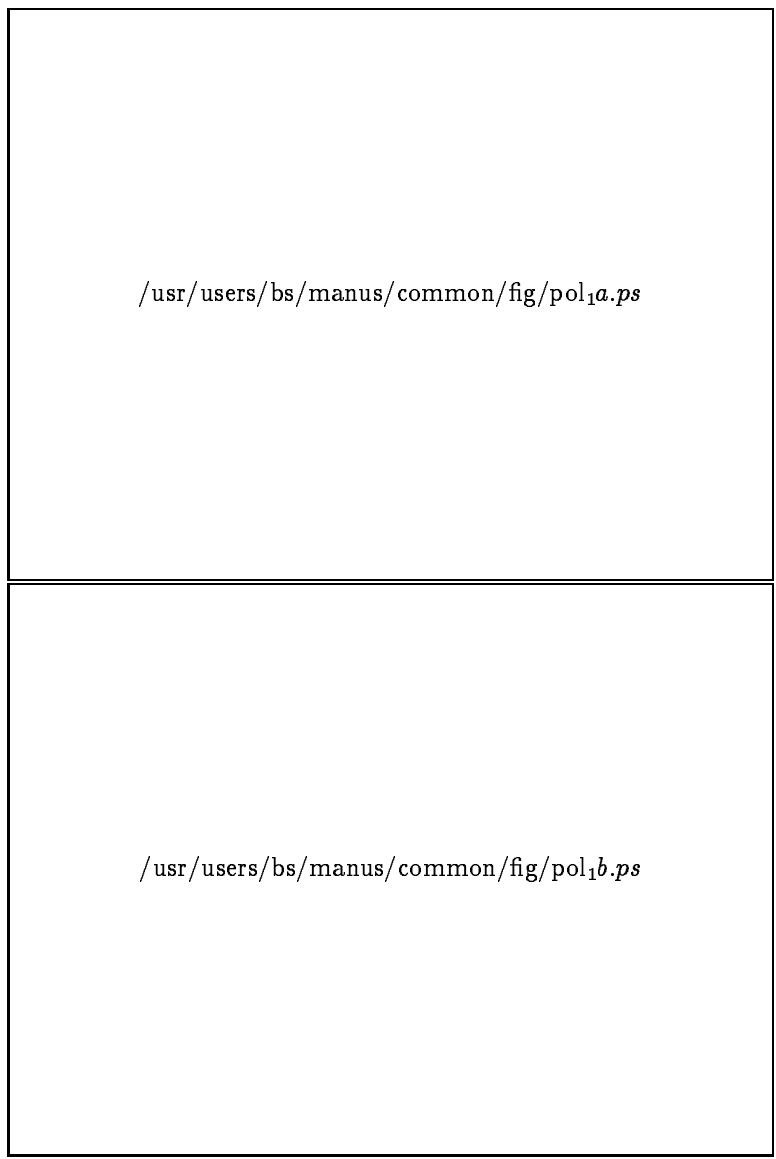

Figure 1: (a) $r_{ee}$ and (b) $r_{mm}$ as functions of $T$ for an unscreened chain with $N = 20$. Filled circles represent MC data, and solid and dashed lines variational results, with $a \neq 0$ and $a = 0$, respectively.

/usr/users/bs/manus/common/fig/pol$_2$a.ps

/usr/users/bs/manus/common/fig/pol$_2$b.ps

Figure 2: (a) $r_{ee}$ and (b) $r_{mm}$ as functions of $T$ for an unscreened chain with $N = 80$. Same notation as in fig. 1.

The remainder of this section will be devoted to comparisons of the variational approach to MC simulation results focusing on (1) energies, in particular the free energy and (2) various configurational measures. The $\mathbf{a}_i = 0$ variational solution will be consistently used.

## 6.1 Free and Internal Energies

In table 1 the variational results for internal Coulombic and Gaussian energies are compared with the MC results; the relative deviations are seen to increase both with increasing $c_s$ and with increasing $N$. For fixed $c_s$ the deviations seem to converge to constant values at large $N$. Hence it should be possible to extract "asymptotic" correction factors from comparisons at moderate $N$, do a variational calculation for a very large $N$, and predict what an MC calculation would give.

|  |  | N | 20 | 40 | 80 | 160 | 320 | 512 |
|---|---|---|---|---|---|---|---|---|
| $c_s=0.0$ | $E_C$ | (V) | 6.20 | 7.58 | 8.80 | 9.94 | 11.0 | 11.7 |
|  |  | (MC) | 5.25 | 6.30 | 7.28 | 8.16 | 8.99 | 9.53 |
|  | $E_G$ | (V) | 6.65 | 7.40 | 8.08 | 8.66 | 9.20 | 9.54 |
|  |  | (MC) | 6.25 | 6.78 | 7.31 | 7.79 | 8.23 | 8.47 |
| $c_s=0.01$ | $E_C$ | (V) | 3.55 | 3.80 | 3.95 | 4.02 | 4.05 | 4.07 |
|  |  | (MC) | 2.70 | 2.83 | 2.89 | 2.91 | 2.93 | 2.93 |
|  | $E_G$ | (V) | 6.20 | 6.63 | 6.88 | 7.00 | 7.07 | 7.10 |
|  |  | (MC) | 5.70 | 6.00 | 6.15 | 6.22 | 6.26 | 6.27 |
| $c_s=0.1$ | $E_C$ | (V) | 1.90 | 2.03 | 2.16 | 2.19 | 2.15 | 2.15 |
|  |  | (MC) | 1.35 | 1.38 | 1.40 | 1.41 | 1.42 | 1.42 |
|  | $E_G$ | (V) | 5.40 | 5.68 | 5.86 | 5.94 | 5.93 | 5.94 |
|  |  | (MC) | 5.00 | 5.15 | 5.26 | 5.29 | 5.32 | 5.32 |
| $c_s=1.0$ | $E_C$ | (V) | 0.65 | 0.70 | 0.74 | 0.75 | 0.76 | 0.76 |
|  |  | (MC) | 0.40 | 0.40 | 0.41 | 0.43 | 0.42 | 0.42 |
|  | $E_G$ | (V) | 4.35 | 4.53 | 4.61 | 4.67 | 4.69 | 4.70 |
|  |  | (MC) | 4.10 | 4.25 | 4.29 | 4.33 | 4.37 | 4.37 |

Table 1: Average internal Coulombic and Gaussian energies per monomer in $kJ/mol \cdot monomer$ for unscreened and screened Coulomb potential. MC and V stands for Monte Carlo and variational calculations respectively. The salt concentration, $c_s$, is given in molar (M).

A strong advantage of the variational approach is the direct access to the free energy, which is much more difficult to obtain in a MC simulation, requiring a cumbersome integration procedure. In order to evaluate the variational results we nevertheless attempt to estimate F(T) from MC data for $N = 20$ using the following procedure. In dimensionless units one has

$$\frac{d(F/T)}{dT} = -\frac{d}{dT}\log \int e^{-E/T}dx = -\frac{1}{T^2}\langle E \rangle \tag{51}$$

Thus, we can define an *excess free energy* with respect to some reference temperature $T_r$ as

$$\Delta F(T) = F(T)/T - F(T_r)/T_r = -\int_{T_r}^{T} \langle E \rangle dT/T^2 \tag{52}$$

which is then accessible in MC by a temperature integration of $\langle E \rangle$. In fig. 3 the excess free energy is shown as a function of $T$, for an $N = 20$ chain with $T_r$ corresponding to 1422 K. As can be seen the variational solutions for $F(T)$ reproduce the extracted MC values very well in a wide temperature

/usr/users/bs/manus/common/fig/pol$_3a$.ps

/usr/users/bs/manus/common/fig/pol$_3b$.ps

Figure 3: The excess free energy $\Delta F(T)$ (see eq. (52)) from the MC data (filled circles) and from the variational approach (solid line with $a \neq 0$ and dotted line with $a = 0$) for **(a)** $c_s = 0.0M$ and **(b)** $c_s = 0.1M$ respectively ($N$=10).

interval. Comparisons for larger $N$ are not feasible due to the indirect cumbersome MC extraction procedure.

We remark that one possibly efficient alternative route to obtain free energies for different degrees of screening at fixed $T$ in MC, would be to perform an integration in the Debye screening length and calculate the incremental excess free energy for a change $\Delta \kappa$. The advantage of such a procedure would be that every point along the integration path corresponds to a physically realistic situation,

| $N$ | | 20 | 40 | 80 | 160 | 320 | 512 | 1024 | 2048 |
|---|---|---|---|---|---|---|---|---|---|
| $r_{mm}$ | V | 13.04 | 13.60 | 14.11 | 14.57 | 14.99 | 15.26 | 15.63 | |
| | MC | 12.56 | 13.01 | 13.43 | 13.81 | 14.17 | 14.38 | 14.68 | 14.99 |
| | diff. | 3.8% | 4.5% | 5.1% | 5.5% | 5.8% | 6.1% | 6.5% | |
| $r_{ee}$ | V | 122 | 277 | 632 | 1425 | 3152 | 5340 | 11478 | |
| | MC | 119 | 269 | 606 | 1347 | 2958 | 4985 | 10651 | 22507 |
| | diff. | 2.5% | 3.9% | 4.3% | 5.8% | 6.6% | 7.1% | 7.8% | |

Table 2: $r_{mm}$ and $r_{ee}$ in Å for unscreened Coulomb potential ($c_s = 0.0M$) as computed with the variational (V) and Monte Carlo (MC) methods. The errors originating from the MC runs are estimated to be $O(0.2\%)$.

while the completely screened chain would serve as a reference state.

## 6.2 The End-End Separation

The end-end separation is a critical measure of the accuracy, being a global quantity with contributions from all bond-bond correlations. Unfortunately, it also turns out to be a complicated quantity from the convergence point of view in standard MC simulations, which means that it will have larger uncertainties than for example the average monomer-monomer separation – this is particularly true for the pure Coulomb chain. Table 2 contains a comparison between variational and simulation results for $r_{mm}$ and $r_{ee}$ using the unscreened Coulomb potential. The result from the variational approach is impressive – the maximal deviation, 7.8% for $N = 1024$, from the MC results is well below the 11% bound discussed above. (Due to memory limitations on the local workstation $N = 2048$ has not been pursued with the variational approach.) This result can be compared to the results from a (spherical) mean field approach [39], which for the same system shows a deviation of 20% already for $N = 100$.

In section 4, $r_{ee}/N$ and $r_{mm}$ were conjectured to vary linearly with $(\log N)^{1/3}$ for large $N$ at zero temperature (cf. eqs. (49, 50)). In fig. 4, this linear dependence is tested on both MC and variational data at room temperature (298K), with a surprisingly good result. This indicates that room temperatures can be considered low for a reasonably long Coulomb chain with the chosen parameters. It is also clearly seen how the variational approximation to $r_{ee}$ exceeds the MC values; asymptotically we expect an 11% discrepancy, as discussed above.

The conjectured zero-temperature scaling results seem to contradict the results by deGennes et al. [40] and Baumgärtner [20] that $r_{ee}$ should scale linearly with $N$ for an unscreened polyelectrolyte. However, the latter result is valid only if the monomer-monomer bonds are rigid, but with elastic bonds as in the present model, eq. (1), swelling is possible and $r_{ee}/N$ increases slowly with $N$ due to the long-range Coulomb repulsion.

Returning to the simulations by Baumgärtner [20], we note that his effective temperature is almost a factor of ten larger than in the present study. Such a high temperature means that the chain behaviour is essentially brownian in character making it numerically difficult to detect the electro-static expansion of the chain in a traditional MC simulation. The rigid monomer-monomer bonds used by Baumgärtner also precludes the extra expansion predicted by eqs. (49,50).

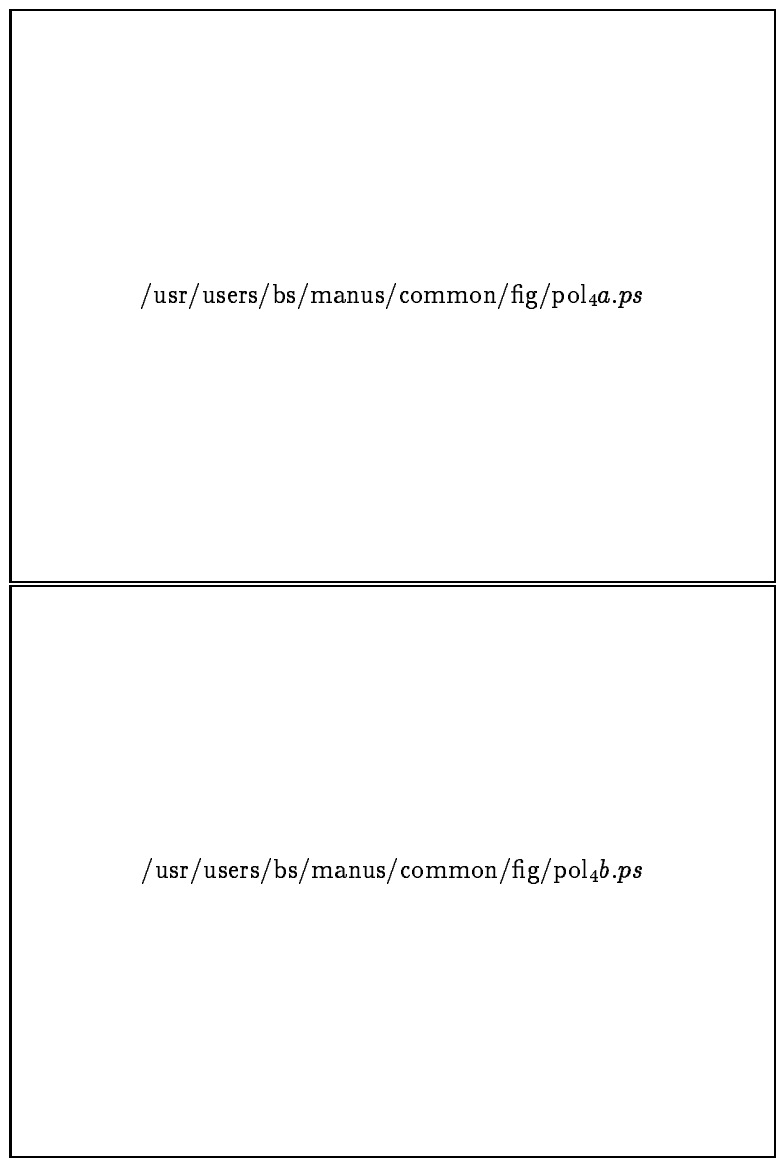

Figure 4: (a) $r_{mm}$ as a function of $(\log N)^{1/3}$. Filled circles represent MC data and solid line the variational results. The dashed line is a linear fit to the MC data. (b) $\log r_{ee}$ as a function of $\log(\log N)$. Filled and open circles represent MC data and variational results respectively. The lines are linear fits.

Table 3 contains the same quantities as table 2, but for a screened Coulomb potential. The agreement detoriates, when a small amount of salt is added. The screening reduces the Coulomb repulsion, which in a sense is the hard part in our variational calculation, and one would naively expect an improved accuracy with a decreased interaction. The opposite result is found and the discrepancy in the end-end separation becomes as large as about 40% with 10 mM of salt and $N = 160$. At sufficiently high salt concentration the agreement improves again, as it should, with the Coulomb repulsion completely screened. With a strong screening the Coulomb potential will have a short range, and for sufficiently large $N$ the chain will be Brownian. This is also reflected in table 3, where the agreement for $r_{ee}$ is worst for intermediate chain lengths and improve again when $N$ increases. The large discrepancy seen e.g. for $c_s$=0.01 M and $N$=160 might seem surprising, considering the excellent agreement found for the pure Coulomb chain, but it reflects the difficulty to properly emulate a short-range potential with a harmonic effective energy [18].

|  |  |  | $N$ | 20 | 40 | 80 | 160 | 320 | 512 |
|---|---|---|---|---|---|---|---|---|---|
| $c_s$=0.01 | $r_{mm}$ | V | | 12.60 | 12.87 | 13.02 | 13.10 | 13.14 | 13.16 |
|  |  | MC | | 12.09 | 12.24 | 12.31 | 12.34 | 12.36 | 12.37 |
|  |  | diff. | | 4.2% | 5.1% | 5.8% | 6.1% | 6.3% | 6.4% |
|  | $r_{ee}$ | V | | 104 | 201 | 377 | 680 | 1188 | 1710 |
|  |  | MC | | 99.6 | 183 | 317 | 521 | 825 | 1110 |
|  |  | diff. | | 4.4% | 9.8% | 19% | 31% | 44% | 54% |
| $c_s$=0.1 | $r_{mm}$ | V | | 11.77 | 11.90 | 11.97 | 12.01 | 12.04 | 12.04 |
|  |  | MC | | 11.30 | 11.35 | 11.39 | 11.38 | 11.40 | 11.40 |
|  |  | diff. | | 4.2% | 4.8% | 5.1% | 5.5% | 5.6% | 5.6% |
|  | $r_{ee}$ | V | | 78.2 | 136 | 231 | 387 | 640 | 895 |
|  |  | MC | | 72.9 | 120 | 192 | 301 | 459 | 622 |
|  |  | diff. | | 7.3% | 13% | 20% | 29% | 39% | 44% |
| $c_s$=1.0 | $r_{mm}$ | V | | 10.57 | 10.69 | 10.69 | 10.69 | 10.70 | 10.70 |
|  |  | MC | | 10.27 | 10.29 | 10.29 | 10.30 | 10.34 | 10.32 |
|  |  | diff. | | 2.9% | 3.9% | 3.9% | 3.8% | 3.5% | 3.7% |
|  | $r_{ee}$ | V | | 55.0 | 86.9 | 137 | 217 | 343 | 468 |
|  |  | MC | | 52.1 | 79.5 | 122 | 182 | 283 | 364 |
|  |  | diff. | | 5.6% | 9.3% | 12% | 19% | 21% | 29% |

Table 3: $r_{mm}$ and $r_{ee}$ in Å for the screened Coulomb potential ($c_s$=0.01M, $c_s$=0.1M and $c_s$=1.0M respectively) as computed with the variational (V) and Monte Carlo (MC) methods. The errors originating from the MC runs are estimated to be $O(0.1\%)$

### 6.3 Monomer-Monomer Separations

The variational results for the average monomer-monomer separation $r_{mm}$ is in excellent agreement with MC results for both the unscreened and screened cases - the largest error seen is of the order of 5% (see tables 2 and 3). As for $r_{ee}$ the variational estimate is always larger than the MC value. This is also true for any single monomer-monomer separation $\langle \mathbf{r}_i^2 \rangle^{1/2}$ as can be seen in fig. 5. The shape of the curves are correctly reproduced by the variational solution and the largest discrepancy is not unexpectedly found in the middle of the chain, which may be explained by a stronger accumulated electrostatic repulsion there. These results can be compared to the mean field solution of ref. [25], in which gradually more and more pair interactions were treated explicitly and withdrawn from the mean field. It was found that only after the inclusion of next-nearest neighbours, leading to lengthy numerical calculations, the curve shape became qualitatively correct.

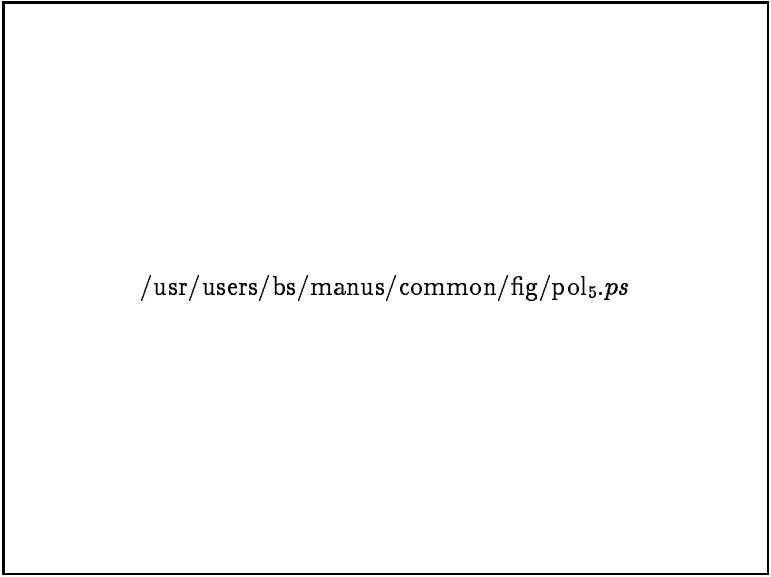

Figure 5: Bond lengths $\langle \mathbf{r}_i^2 \rangle^{1/2}$ along an $N = 40$ chain. Salt concentrations $c_s$ from top to bottom are 0 M, 0.01M, 0.1M and 1 M respectively. Solid line represents variational results and dashed line Monte Carlo data.

We discussed above the zero-temperature scaling for an unscreened chain, $r_{mm} \propto (\log N)^{1/3}$. When salt is added $r_{mm}$ increases much more slowly with chain length and for $c_s = 1$ M, when $r_{mm} \gg \kappa^{-1}$, it is essentially independent of chain length. One also notes that the individual bond lengths $\langle r_i^2 \rangle^{1/2}$ do not vary in the central part of the chain when salt is present (see fig. 5).

Furthermore, for an unscreened chain, following eq. (48), we expect at $T = 0$ the individual bond-lengths, raised to the power three, to vary linearly with $u = \log[s(1-s)]$, where $s = i/N$. In fig. 6a it is shown that this is approximately true, more so for the variational solution than for the MC results. The lower curve in fig. 6a, obtained with a screened Coulomb potential, shows a qualitatively different behaviour. Fig. 6b contains a similar graph for different $N$.

The $T = 0$ scaling relation for individual bonds, eq. (48), has the peculiar consequence that the length of a bond at the end of the chain becomes independent of $N$. This can be seen by rewriting the scaling relation as

$$\langle \mathbf{r}_i^2 \rangle^{1/2} \approx \{\log N + \log[s(1-s)]\}^{1/3} \approx [\log i]^{1/3} \qquad (53)$$

where the last expression holds for small $i$. Eq. (53) also holds for the MC results, where we find the first few bond lengths to be independent of $N$.

/usr/users/bs/manus/common/fig/pol$_6$a.ps

/usr/users/bs/manus/common/fig/pol$_6$b.ps

Figure 6: The cubed bond length $\langle \mathbf{r}_i^2 \rangle^{3/2}$ as a function of $\log(s(1-s))$, with $s = i/N$ the relative monomer position. (**a**) Upper line shows variational results and symbols MC data for $N = 40$, no screening. The zero temperature scaling corresponds to a straight line. The lower line shows MC data for a screened chain ($c_s = 0.01$ M) for comparison. (**b**) Variational results for unscreened chains of varying size $N$.

## 6.4 Angular Correlations

In order to further test the variational solutions, we also have calculated angular correlations between bonds,

$$C_{i,j} = \frac{\langle \mathbf{r}_i \cdot \mathbf{r}_j \rangle}{\sqrt{\langle \mathbf{r}_i^2 \rangle \langle \mathbf{r}_j^2 \rangle}} \tag{54}$$

which roughly gives the average of the cosine between bonds. In fig. 7a, variational and MC data is shown for the neighbor correlation $C_{i,i+1}$. It is seen that the variational Ansatz consistently overestimates the angle, more so in the presence of salt than without. Comparing figs. 5 and 7a, we find that the above discussed discrepancy between the MC and variational results for the end-end separation seems to be due to differences in angular correlations as well as in bond lengths. For the unscreened case the two sources seem to be of comparable magnitude, while for the screened case the angular correlations seem to be larger and the main cause of discrepancy.

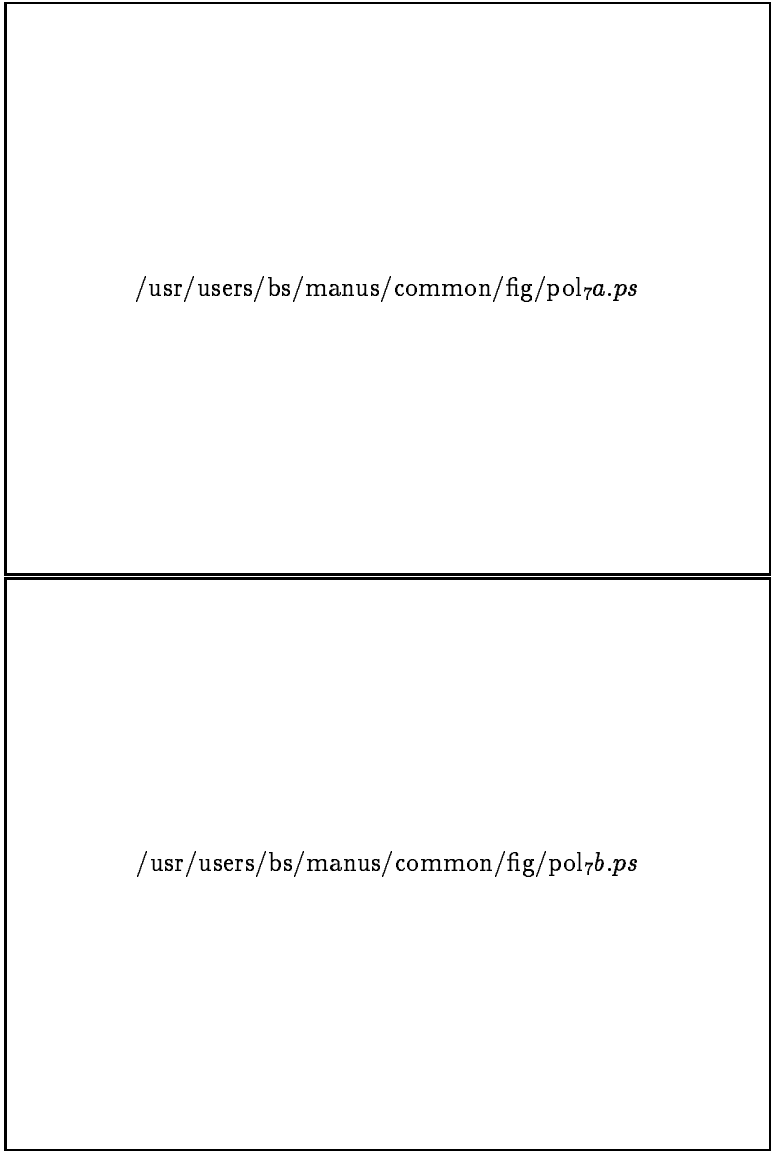

Figure 7: Angular correlations (a) between neighbouring bonds, and (b) between the first and successive bonds. Solid lines represent variational and dashed lines MC results. The upper pair of curves are for an unscreened chain, while the lower pair is for $c_s = 0.01$ M. ($N = 80$.)

Fig. 7b shows a more global angular correlation $C_{1,i}$ between the first and all successive bonds. The variational results for this quantity are in excellent agreement with MC data, both in the screened and in the unscreened chain.

# 7   Summary and Outlook

A deterministic variational scheme for discrete representations of polymer chains has been presented, where the true bond and Coulombic potentials are approximated with a trial isotropic harmonic energy. The variational parameters obey matrix equations, for which a very effective iterative solution scheme has been developed – the computational demand is $N^3$.

The high and low T properties of the variational approach has been analyzed with encouraging results. Also, the approach is shown to obey the relevant virial identities.

In contrast to MC simulations, the free energy is directly accessible with the variational method.

When confronting the results from the method with those from MC simulations, very good agreement is found for configurational quantities in the case of an unscreened Coulomb interaction (the error is within 11 %).

In the screened case the method does not reproduce the MC results equally well although the qualitative picture of conformational properties is there. We attribute this problem to the difficulty for a Gaussian to emulate short range interactions.

Recently, MC simulations were pursued for titrating Coulomb chains [41]. For such systems the Coulomb potential of eq. (1) is modified to

$$\frac{q^2}{4\pi\epsilon_r\epsilon_0} \sum_{i<j} \frac{s_i s_j}{|\tilde{\mathbf{x}}_{ij}|} \tag{55}$$

where the binary variables $s_i$ are either 1 or 0 depending whether monomer $i$ is charged or not. Thus minimizing E now also includes a combinatorial problem – deciding where the charges should be located. Variational techniques related to the ones used in this paper have been successfully used in pure combinatorial optimization problems [42], where again tedious stochastic procedures are replaced by a set of deterministic equations. Along similar lines, the approach of this paper can be modified to allow for a variational treatment also of the titrating problem.

The variational approach is also directly applicable to more general topologies – bifurcations pose no problems. Proteins could also be treated in this way provided the traditional Lennard-Jones potentials are replaced by forms that are less singular at the origin.

## Acknowledgements:

We wish to thank Anders Irbäck for providing us with some of the MC data points. One of us (BJ) wants to acknowledge many stimulating discussions with R. Podgornik during the course of this work.

# Appendix A. The Variational Approach – Generalities

In this appendix we discuss the generic variational approach that is used in this paper for the particular problem of a Coulomb chain. We consider a generic system, with dynamical variables $x$ in some multi-dimensional state-space, and assume that a real energy function $E(x)$ is given.

For an arbitrary probability distribution $P(x)$ in an arbitrary state space, the *free energy* with respect to an energy $E(x)$ is generally defined as

$$\hat{F} = \langle E \rangle - TS \tag{A1}$$

where $S$ is the *entropy*,

$$S = -\langle \log P \rangle \tag{A2}$$

and expectation values are defined with respect to $P$. Writing $P(x)$ as

$$P(x) = \frac{1}{Z} \exp(-E_V(x)/T) \tag{A3}$$

with

$$Z = \int dx \; \exp(-E_V(x)/T) \tag{A4}$$

the free energy can be written as

$$\hat{F} = -T \log Z + \langle E - E_V \rangle \tag{A5}$$

Note that

$$
\begin{aligned}
\exp(-\hat{F}/T) &= Z \exp\left\langle \frac{E_V - E}{T} \right\rangle \\
&\leq Z \left\langle \exp\left(\frac{E_V - E}{T}\right) \right\rangle = \int dx \; \exp(-E/T) \\
&= \exp(-\hat{F}/T)|_{E_V = E}
\end{aligned}
\tag{A6}
$$

where the inequality is due to the convexity of the exponential function. Thus, $\hat{F}$ is bounded from below by its value for $E_V = E$, corresponding to the proper Boltzmann distribution, $P(x) \propto \exp(-E(x)/T)$.

The variation of $\hat{F}$ due to a variation $\delta E_V$ is given by

$$
\begin{aligned}
\delta \hat{F} &= -T \left( \langle \delta E_V (E - E_V) \rangle - \langle \delta E_V \rangle \langle E - E_V \rangle \right) \\
&\equiv -T \langle \delta E_V (E - E_V) \rangle_C
\end{aligned}
\tag{A7}
$$

where $\langle ab \rangle_C$ stands for the connected expectation value (cumulant), $\langle ab \rangle - \langle a \rangle \langle b \rangle$.

The idea of the *variational approach* is to choose a suitable simple Ansatz for the variational energy $E_V$, with a set of adjustable parameters $\alpha_i$, $i = 1, \ldots, N_p$, the values of which are to be chosen so as to minimize the variational free energy. Demanding the vanishing of the variation of the free energy due to variations in the parameters $\alpha_i$ then leads to the general equations for an extremum:

$$\left\langle \frac{\partial E_V}{\partial \alpha_i} (E - E_V) \right\rangle_C^V = 0, \quad i = 1, \ldots, N_p \tag{A8}$$

where $\langle\rangle^V$ denotes an expectation value based on the variational Boltzmann distribution. This determines the optimal values of the parameters. Exact expectation values are then approximated by the corresponding variational ones.

# Appendix B. The Variational Free Energy for the Chain

In this appendix we derive the expressions for the variational free energy (eqs. (26, 31)), for the unscreened as well as the screened Coulomb chain, with or without translational parameters in the variational Ansatz.

## Unscreened Coulomb Chain

For the specific case of the Coulomb chain of length $N$, the energy amounts to

$$E = \frac{1}{2} \sum_i \mathbf{r}_i^2 + \sum_\sigma \frac{1}{r_\sigma} \tag{B1}$$

where $\sigma$ is a contiguous subchain.

**Gaussian Parameters Only**

We consider first a pure Gaussian variational Boltzmann distribution, corresponding to

$$E_V/T = \frac{1}{2} \sum_{i,j} G^{-1}_{ij} \mathbf{r}_i \cdot \mathbf{r}_j \tag{B2}$$

The parameter matrix $G$ is forced to be symmetric and positive-definite by expressing it as

$$G_{ij} = \mathbf{z}_i \cdot \mathbf{z}_j = \sum_{\mu=1}^{N-1} z_{i\mu} z_{j\mu} \tag{B3}$$

The general expression for the variational free energy is

$$\hat{F} = -T \log Z_V - \langle E_V \rangle_V + \langle E \rangle_V \tag{B4}$$

where the expectation values are with respect to the normalized variational Boltzmann distribution $\exp(-E_V/T)/Z_V$. The first two terms are trivial to compute. The first is

$$-T \log Z_V = \frac{3T}{2} \log \det G^{-1} \equiv -3T \log \det z \tag{B5}$$

apart from a trivial constant that can be neglected, as can the second term,

$$-\langle E_V \rangle_V = -\frac{3}{2}(N-1)T \tag{B6}$$

The last term,

$$\langle E \rangle_V = \frac{1}{2} \sum_i \langle \mathbf{r}_i^2 \rangle_V + \sum_\sigma \left\langle \frac{1}{r_\sigma} \right\rangle_V \tag{B7}$$

consists in a sum of terms, each amounting to the variational expectation value of a simple function of a Gaussian vector variable $\mathbf{r}_\sigma$, of which $\mathbf{r}_i$ is a special case. Its probability distribution is given by

$$P(\mathbf{r}_\sigma) \propto \exp\left(-\frac{\mathbf{r}_\sigma^2}{2\mathbf{z}_\sigma^2}\right) \tag{B8}$$

with

$$\mathbf{z}_\sigma \equiv \sum_{i \in \sigma} \mathbf{z}_i \tag{B9}$$

Thus, we have

$$\langle \mathbf{r}_i^2 \rangle_V = 3\mathbf{z}_i^2 \tag{B10}$$

and

$$\left\langle \frac{1}{r_\sigma} \right\rangle_V = \sqrt{\frac{2}{\pi}} \frac{1}{z_\sigma} \equiv U_1^C(z_\sigma) \tag{B11}$$

Summing up, the variational free energy takes the form

$$\hat{F} = -3T \log \det z + \frac{3}{2} \sum_i \mathbf{z}_i^2 + \sum_\sigma U_1^C(z_\sigma) \tag{B12}$$

to be minimized with respect to the variational parameters $\mathbf{z}_i$.

**Gaussian and Translational Parameters**

For the more general variational Ansatz with additional translational parameters $\mathbf{a}_i$,

$$E_V/T = \frac{1}{2} \sum_{i,j} G_{ij}^{-1} (\mathbf{r}_i - \mathbf{a}_i) \cdot (\mathbf{r}_j - \mathbf{a}_j) \tag{B13}$$

the main difference is a translation of the individual probability distributions of eq. (B8), which now read

$$P(\mathbf{r}_\sigma) \propto \exp\left(-\frac{(\mathbf{r}_\sigma - \mathbf{a}_\sigma)^2}{2\mathbf{z}_\sigma^2}\right) \tag{B14}$$

with

$$\mathbf{a}_\sigma \equiv \sum_{i \in \sigma} \mathbf{a}_i \tag{B15}$$

This gives

$$\langle \mathbf{r}_i^2 \rangle_V = 3\mathbf{z}_i^2 + \mathbf{a}_i^2 \tag{B16}$$

and

$$\left\langle \frac{1}{r_\sigma} \right\rangle_V = \frac{1}{a_\sigma} \operatorname{erf}\left(\frac{a_\sigma}{\sqrt{2}z_\sigma}\right) \equiv U_2^C(z_\sigma, a_\sigma) \tag{B17}$$

The variational free energy becomes

$$\hat{F} = -3T \log \det z + \frac{1}{2} \sum_i (3\mathbf{z}_i^2 + \mathbf{a}_i^2) + \sum_\sigma U_2^C(z_\sigma, a_\sigma) \tag{B18}$$

## Screened Coulomb Chain

Next we consider the Debye-screened version of the Coulumb chain, with the energy

$$E = \frac{1}{2} \sum_i \mathbf{r}_i^2 + \sum_\sigma \frac{\exp(-\kappa r_\sigma)}{r_\sigma} \tag{B19}$$

### Gaussian Parameters Only

For the variational Ansatz with only Gaussian parameters, eq. (B2), we need eq. (B10) and the expectation value

$$\left\langle \frac{\exp(-\kappa r_\sigma)}{r_\sigma} \right\rangle_V = \sqrt{\frac{2}{\pi}} \frac{1}{z_\sigma} - \kappa \exp\left(\frac{\kappa^2 z_\sigma^2}{2}\right) \operatorname{erfc}\left(\frac{\kappa z_\sigma}{\sqrt{2}}\right) \equiv U_1^D(z_\sigma) \tag{B20}$$

The variational free energy then reads

$$\hat{F} = -3T \log \det z + \frac{3}{2} \sum_i \mathbf{z}_i^2 + \sum_\sigma U_1^D(z_\sigma) \tag{B21}$$

### Gaussian and Translational Parameters

Finally, if for the screened Coulomb chain, eq. (B19), also translational parameters are used in $E_V$, eq. (B13), we will need the following result in addition to eq. (B16),

$$\left\langle \frac{\exp(-\kappa r_\sigma)}{r_\sigma} \right\rangle_V = \frac{\exp(\kappa^2 z_\sigma^2/2)}{2a_\sigma} \left[ \exp(-\kappa a_\sigma) \operatorname{erfc}\left(\frac{\kappa z_\sigma^2 - a_\sigma}{\sqrt{2} z_\sigma}\right) - \exp(\kappa a_\sigma) \operatorname{erfc}\left(\frac{\kappa z_\sigma^2 + a_\sigma}{\sqrt{2} z_\sigma}\right) \right]$$
$$\equiv U_2^D(z_\sigma, a_\sigma) \tag{B22}$$

The variational free energy will read

$$\hat{F} = -3T \log \det z + \frac{1}{2} \sum_i (3\mathbf{z}_i^2 + \mathbf{a}_i^2) + \sum_\sigma U_2^D(z_\sigma, a_\sigma) \tag{B23}$$

# Appendix C. The Virial Identity

In this appendix we derive the virial identities and show that these are respected by the variational approach.

## Exact Virial Identity

For any system described by a Boltzmann distribution

$$P(\mathbf{x}) = \frac{1}{Z}\exp(-E(\mathbf{x})/T) \tag{C1}$$

with $\mathbf{x} \in \mathcal{R}^D$, and $E$ rising as a power for large $|\mathbf{x}|$, we will have

$$\frac{1}{Z}\int \nabla \cdot (\mathbf{f}(\mathbf{x})\exp(-E/T))dx = 0 \tag{C2}$$

for e.g. any polynomial $\mathbf{f}$, due to the integrand being an exact divergence. This is equivalent to

$$T\langle \nabla \cdot \mathbf{f}\rangle = \langle \mathbf{f} \cdot \nabla E\rangle \tag{C3}$$

Thus, by varying $\mathbf{f}$, we can obtain an infinite set of identities for the system.

The *virial identity* results from the particular choice $\mathbf{f} = \mathbf{x}$; in its general form it reads

$$\langle \mathbf{x} \cdot \nabla E\rangle = TD \tag{C4}$$

where $D$ is the dimension of $\mathbf{x}$-space ($\mathbf{x} \cdot \nabla$ is the *scaling operator*).

This is particularly useful if the energy $E$ is given by a sum of terms $E_a$ homogeneous in $\mathbf{x}$,

$$\mathbf{x} \cdot \nabla E_a = \lambda_a E_a \tag{C5}$$

in which case the virial identity takes the simple form

$$\sum_a \lambda_a \langle E_a\rangle = TD \tag{C6}$$

This applies e.g. to the case of the unscreened polyelectrolyte. There the scaling operator is given, in relative coordinates, by $\sum_i \mathbf{r}_i \cdot \nabla_{\mathbf{r}_i}$, and we have $\lambda_G = 2$ and $\lambda_C = -1$; the virial identity thus reads

$$2\langle E_G\rangle - \langle E_C\rangle = 3(N-1)T \tag{C7}$$

## Variational Virial Identity

The virial identity is preserved by the variational approach under certain conditions, to be specified below. For the generic system above, minimizing the free energy

$$\hat{F} = F_V + \langle E - E_V\rangle_V \tag{C8}$$

w.r.t. the parameters $\alpha_i$ of a variational energy $E_V$, leads to (cf. eq. (A8) in Appendix A)

$$-T\frac{\partial \hat{F}}{\partial \alpha_i} \equiv \left\langle (E - E_V)\frac{\partial E_V}{\partial \alpha_i} \right\rangle_C^V = 0 \tag{C9}$$

Now, choosing $\mathbf{f} = \mathbf{x}(E - E_V)$ in eq. (C3) with the variational Boltzmann distribution, we have

$$DT\langle E - E_V\rangle_V + T\langle \mathbf{x} \cdot \nabla(E - E_V)\rangle_V = \langle (E - E_V)\mathbf{x} \cdot \nabla E_V\rangle_V \tag{C10}$$

On the other hand, since the virial theorem holds for $E_V$,

$$DT = \langle \mathbf{x} \cdot \nabla E_V\rangle_V \tag{C11}$$

Substituting this into eq. (C10), we obtain

$$T\langle \mathbf{x} \cdot \nabla(E - E_V)\rangle_V = \langle (E - E_V)\mathbf{x} \cdot \nabla E_V\rangle_C^V \tag{C12}$$

If the set of parameters $\alpha_i$ of $E_V$ is such (and this is the crucial condition), that the scaling operation on $E_V$ can be written in terms of derivatives with respect to the parameters, i.e. if

$$\mathbf{x} \cdot \nabla E_V = \sum_i G_i(\alpha)\frac{\partial E_V}{\partial \alpha_i} \tag{C13}$$

then the righthand side of eq. (C12) vanishes at the minimum, due to eq. (C9), and we are left with

$$\langle \mathbf{x} \cdot \nabla E\rangle_V = \langle \mathbf{x} \cdot \nabla E_V\rangle_V = DT \tag{C14}$$

which is what we desired.

Note that the derivation only relies on a local extremum of the free energy. Thus, for the polymer, the virial identity, eq. (C7), is respected by both the rigid ($\mathbf{a} \neq 0$) and the purely fluctuating ($\mathbf{a} = 0$) solutions.

# Appendix D. High and Low T Expansions

## High T expansions

**Exact results**

At high $T$, the chain size will be large, and accordingly, the Gaussian term will dominate over the interaction $V$ in the energy expression,

$$E = E_G + V = \frac{1}{2} \sum_i r_i^2 + \sum_\sigma v(r_\sigma) \tag{D1}$$

with $\sigma$ summed over contiguous subchains.

It is then natural to attempt an expansion in the perturbation $V$. For an arbitrary expectation value, we have the perturbative expansion

$$\langle f \rangle = \langle f \rangle^0 - \frac{1}{T} \langle fV \rangle_C^0 + \frac{1}{2T^2} \langle fVV \rangle_C^0 - \ldots \tag{D2}$$

where $\langle \; \rangle_C^0$ refers to connected expectation values (cumulants) in the unperturbed Boltzmann distribution. Due to the singular behaviour of $v(r)$ for small $r$ in the (screened or unscreened) Coulomb case, only the first few terms will be finite. This indicates that expectation values cannot be expanded in a pure power series in $T$, and that logarithmic corrections will occur after the first finite terms.

We are interested in quadratic expectation values of the type $\langle \mathbf{r}_i \cdot \mathbf{r}_j \rangle$. These can be combined to give e.g. the rms end-to-end distance $r_{ee}$, the gyration radius, and the Gaussian energy $\langle E_G \rangle$ (and thereby, in the pure Coulomb case, the interaction energy $\langle E_C \rangle$ by the virial identity).

For the pure Coulomb chain, $v(r) = 1/r$, we get the perturbative expansion

$$\langle \mathbf{r}_i \cdot \mathbf{r}_j \rangle = 3T\delta_{ij} + \sqrt{\frac{2}{\pi T}} \sum_{\sigma \ni i,j} L_\sigma^{-3/2} + O(T^{-2}) \tag{D3}$$

where $\sigma$ denotes a contiguous sub-chain containing the $i$th and the $j$th bond, and $L_\sigma$ its total number of bonds. This leads to

$$\langle E_G \rangle = 3(N-1)T/2 + \sqrt{\frac{1}{2\pi T}} \sum_\sigma L_\sigma^{-1/2} + O(T^{-2}) \tag{D4}$$

and

$$r_{ee}^2 = 3(N-1)T + \sqrt{\frac{2}{\pi T}} \sum_\sigma L_\sigma^{1/2} + O(T^{-2}) \tag{D5}$$

Similar results are obtained for the case of a screened Coulomb interaction, $v(r) = e^{-\kappa r}/r$, where the expansion of a quadratic expectation value gives

$$\langle \mathbf{r}_i \cdot \mathbf{r}_j \rangle = 3T\delta_{ij} + \frac{3}{\kappa^2 T} \sqrt{\frac{2}{\pi T}} \sum_{\sigma \ni i,j} L_\sigma^{-5/2} + O(T^{-3}) \tag{D6}$$

**Variational results**

The corresponding variational results can also be expanded at high $T$ (where $\mathbf{a}_i = 0$), by expanding the variational solution $z_{i\mu}$ around the unperturbed value, which can be chosen as $\sqrt{T}\delta_{i\mu}$. The variational approximation to a quadratic expectation value, $\langle \mathbf{r}_i \cdot \mathbf{r}_j \rangle_V = 3\mathbf{z}_i \cdot \mathbf{z}_j$, can then easily be expanded. The results thus obtained reproduce the exact results, eqs. (D3,D6), correctly to the order shown.

We conclude that, independently of screening, the high $T$ variational results are correct to next-to-leading order for $E$ and $E_G$, and to leading order for $E_C$.

## Low T expansions

Here we will treat only the pure Coulomb case in detail; most of the discussion applies also to the screened case.

**Exact results**

At low $T$, expectation values can be expanded around the configuration that minimizes the energy. This expansion is slightly complicated by the the rotational degeneracy of the minimum. The results can be expressed in terms of the classical configuration, which is given by a straight line configuration.

Let $b_i$ be the the bond-lengths at the energy minimum. These have to be computed numerically, by solving the equation

$$b_i = \sum_{\sigma \ni i} \frac{1}{b_\sigma^2} \tag{D7}$$

where $b_\sigma$ is the length of a subchain containing the $i$th bond. Note that the above equation can be written as a matrix equation:

$$b_i = \sum_j b_j \sum_{\sigma \ni i,j} \frac{1}{b_\sigma^3} = \sum_j B_{ij} b_j \tag{D8}$$

Thus, $b$ is an eigenvector of the matrix $B$ with a unit eigenvalue. Similarly, we can define a whole series of tensors:

$$E_C = \sum_\sigma \frac{1}{b_\sigma} \tag{D9}$$

$$A_i = \sum_{\sigma \ni i} \frac{1}{b_\sigma^2} \equiv b_i \tag{D10}$$

$$B_{ij} = \sum_{\sigma \ni i,j} \frac{1}{b_\sigma^3} \tag{D11}$$

$$C_{ijk} = \sum_{\sigma \ni i,j,k} \frac{1}{b_\sigma^4} \tag{D12}$$

They are all symmetric, and contracting either with $b_i$ gives the tensor of rank one less.

In addition, we need two more matrices, related to $B$,

$$U = (1+2B)^{-1} \qquad (D13)$$
$$V = P(1-B)^{-1}P \qquad (D14)$$

where $P$ denotes the projection matrix onto the subspace orthogonal to $b$, which is deleted by $1-B$.

In terms of these tensors, we have the quadratic expectation-values at low $T$:

$$\langle \mathbf{r}_i \cdot \mathbf{r}_j \rangle = b_i b_j + T \left( U_{ij} + 2V_{ij} + \frac{4 b_i b_j}{3 \sum_k b_k^2} + 3 \sum_{klm} C_{klm}(b_i U_{jk} + b_j U_{ik})(U_{lm} - V_{lm}) \right) + O(T^2) \quad (D15)$$

where the first two terms of the $T$ coefficient are the naive contributions from the longitudinal and transverse fluctuations. The rest are corrections due to the rotational degeneracy of the $T=0$ configuration, which is also responsible for the transverse zero-modes (of $1-B$).

From this we obtain e.g the average Gaussian energy,

$$\langle E_G \rangle = 1/3 E_0 + T(3N/2 - 11/6) + O(T^2)q \qquad (D16)$$

from which, using the virial identity, we obtain

$$\langle E_C \rangle = 2/3 E_0 - 2T/3 + O(T^2) \qquad (D17)$$

and

$$\langle E \rangle = E_0 + T(3N - 5)/2 + O(T^2) \qquad (D18)$$

where $E_0 = 3/2 \sum_i b_i^2$ is the exact energy at $T = 0$. In the last equation, the $T$-coefficient is, as it should, half the number of degrees of freedom, not counting the two rotational zero-modes (and the three translational ones already removed).

In the screened case, similar results can be obtained. In particular, eq. (D18) remains valid, though with a different $E_0$.

**Variational results**

Similarly, the variational results can be expanded at low $T$. We have to distinguish between the two different solutions.

For the *purely fluctuating* solution with $\mathbf{a}_i = 0$, the variational free energy is, at $T = 0$,

$$\hat{F}_0 \equiv \langle E \rangle_V = \frac{3}{2} \sum_i \mathbf{z}_i^2 + \sqrt{\frac{2}{\pi}} \sum_\sigma \frac{1}{z_\sigma} \qquad (D19)$$

This is obviously just the energy of an $(N-1)$-dimensional version of the chain, with modified coefficients. It is minimized by the aligned configuration

$$\mathbf{z}_i = \left( \frac{2}{9\pi} \right)^{1/6} b_i \hat{\mathbf{n}} \qquad (D20)$$

where $b_i$ are given by eq. (D7), and $\hat{\mathbf{n}}$ is some unit vector in $N-1$ dimensions. For small but finite $T$, we have to add the entropy term, $-3T\log\det\{\mathbf{z}\}$, to $\hat{F}_0$. This forces the configuration out of alignment. The resulting configuration can be obtained as a low-$T$ expansion around the $T=0$ solution. The first correction to $\mathbf{z}$ will be of order $\sqrt{T}$, and since the $T=0$ $\mathbf{z}_i$ are aligned, the matrix inverse $\mathbf{w}_i$ diverges - it will go like $1/\sqrt{T}$.

For the total energy, the leading correction can be obtained as follows. The equation to solve is

$$3T\mathbf{w}_i = \nabla_i \hat{F}_0(\mathbf{z}) \tag{D21}$$

For the $T=0$ solution, $\nabla_i \hat{F}_0(\mathbf{z}_0) = 0$. Thus, to lowest order,

$$3T\mathbf{w}_i = \nabla_i \sum_j \nabla_j \hat{F}_0(\mathbf{z}_0) \cdot d\mathbf{z}_j \tag{D22}$$

The leading energy correction will be

$$d\hat{F}_0 = \frac{1}{2}\sum_{ij} \nabla_i \nabla_j \hat{F}_0(\mathbf{z}_0) d\mathbf{z}_j d\mathbf{z}_i = \frac{3T}{2}\sum_i \mathbf{w}_i \cdot d\mathbf{z}_i \tag{D23}$$

Now, $d\mathbf{z}_i$ consists of an aligned part and a transverse part, both $\propto \sqrt{T}$, while for $\mathbf{w}_i$ the aligned part is $\propto 1$, while the transverse part is $\propto 1/\sqrt{T}$. Thus, the leading contribution to $d\hat{F}_0$ comes from the transverse part. Taking the trace of the tranverse part of the identity $\mathbf{w}_i \cdot \mathbf{z}_j = \delta_{ij}$, and noting that the transverse part of $\mathbf{z}$ sits entirely in $d\mathbf{z}$, we have to leading order

$$d\hat{F}_0 = \frac{3T(N-2)}{2} \equiv d\langle E\rangle_V \tag{D24}$$

Using the virial identity, which holds also for the variational expectation values, we get to first order in $T$:

$$\langle E\rangle_V = (\frac{6}{\pi})^{1/3} E_0 + \frac{3T(N-2)}{2} + O(T^2) \tag{D25}$$

$$\langle E_G\rangle_V = \frac{1}{3}(\frac{6}{\pi})^{1/3} E_0 + \frac{T(3N-4)}{2} + O(T^2) \tag{D26}$$

$$\langle E_C\rangle_V = \frac{2}{3}(\frac{6}{\pi})^{1/3} E_0 - T + O(T^2) \tag{D27}$$

where $E_0$ is the exact $T=0$ energy. Note that already the zero-order results are off by a factor $(\frac{6}{\pi})^{1/3} \approx 1.24$, but that the correction to $E$ is correct in the high $N$ limit.

Similar results are obtained for the screened Coulomb chain. Most of the general analysis leading to eq.(D24) still holds, and we have e.g.

$$\langle E\rangle_V = E_0' + \frac{3T(N-2)}{2} + O(T^2) \tag{D28}$$

with $E_0' \neq E_0$.

For the *symmetry-broken* $\mathbf{a}_i \neq 0$ solution, the $T=0$ configuration is instead given by

$$\mathbf{a}_i = b_i \hat{\mathbf{n}} \ , \ \mathbf{z}_i = 0 \tag{D29}$$

with $b_i$ as in eq. (D7), and $\hat{\mathbf{n}}$ an arbitrary unit vector in $\mathcal{R}^3$. The small $T$ corrections are obtained from expressing $\hat{F}$ as

$$\hat{F} = -3T \log \det\{\mathbf{z}\} + \langle E(\mathbf{a}_i + \sum_\mu z_{i\mu} \mathbf{J}_\mu) \rangle_J \tag{D30}$$

where $\mathbf{J}_\mu$ are uncorrelated standard Gaussian noise variables. Expanding in $\mathbf{z}$, we obtain

$$\hat{F} = -3T \log \det\{\mathbf{z}\} + E(\mathbf{a}) + \frac{1}{2} \sum_{ij} \mathbf{z}_i \cdot \mathbf{z}_j \nabla_i \cdot \nabla_j E(\mathbf{a}) + \ldots \tag{D31}$$

Because the Coulomb term satisfies Laplace' equation, this is just (provided $\mathbf{a} \neq 0$)

$$\hat{F} = -3T \log \det\{\mathbf{z}\} + E(\mathbf{a}) + \frac{3}{2} \sum_i \mathbf{z}_i \mathbf{z}_i \tag{D32}$$

and the variational free energy separates in $\mathbf{z}$ and $\mathbf{a}$ for small $\mathbf{z}$. Thus, minimum is obtained for

$$\mathbf{a}_i = \hat{\mathbf{n}} b_i \tag{D33}$$

and $\mathbf{z}$ satisfies

$$-3T \mathbf{w}_i + 3\mathbf{z}_i = 0 \tag{D34}$$

which means

$$\mathbf{z}_i \cdot \mathbf{z}_j = T \delta_{ij} \tag{D35}$$

The energies become

$$\langle E \rangle_V = E_0 + \frac{3T(N-1)}{2} \tag{D36}$$

$$\langle E_G \rangle_V = \frac{1}{3} E_0 + \frac{3T(N-1)}{2} \tag{D37}$$

$$\langle E_C \rangle_V = \frac{2}{3} E_0 \tag{D38}$$

which is correct to lowest order, and for large $N$ qualitatively correct to first order (except for $E_C$).

Again, the screened chain lead to similar results; in particular, the $T = 0$ energies will be the correct ones.

# Appendix E. Zero Temperature Scaling Properties

The $T = 0$ configuration of a pure Coulombic chain cannot be obtained analytically, but must be computed numerically. However, an approximate calculation can be done. The equation for the bond lengths $b_i$ in the elongated ground state configuration is given by eq. (D7):

$$b_i = \sum_{\sigma \ni i} \frac{1}{b_\sigma^2} \tag{E1}$$

By assuming that the bond length is locally approximately constant, this can be approximated by

$$b_i \approx \frac{1}{b_i^2} \sum_{k=1}^{i} \sum_{l=i}^{N-1} (l - k + 1)^{-2} \tag{E2}$$

This can be rewritten as

$$b_i^3 \approx \sum_{l=1}^{N-1} (-l + \min(l, N - i) + \min(l, i))/l^2 \tag{E3}$$

This in turn can be approximated by an integral, leading to

$$b_i \approx \left[ \log \left( \text{const} \frac{i(N - i)}{N} \right) \right]^{1/3} \tag{E4}$$

Defining $s = i/N$, this amounts to

$$b_i \approx (\log(\text{const } Ns(1 - s)))^{1/3} \tag{E5}$$

Eqs. (E4,E5) give a quite accurate picture of the variation of the bond lengths at $T = 0$. They also imply, that the typical ground state bond length should swell roughly as $(\log N)^{1/3}$ for large $N$.